# Two dimensional crystals in three dimensions: electronic decoupling of single-layered platelets in colloidal nanoparticles.

*Roman Kempt, Agnieszka Kuc, Jae Hyo Han, Jinwoo Cheon, Thomas Heine\**

Roman Kempt
Wilhelm-Ostwald-Institut für Physikalische und Theoretische Chemie, Universität Leipzig, Linnéstr. 2, 04103 Leipzig, Germany

Agnieszka Kuc
Helmholtz-Zentrum Dresden-Rossendorf, Abteilung Ressourcenökologie, Forschungsstelle Leipzig, Permoserstr. 15, 04318 Leipzig, Germany
Wilhelm-Ostwald-Institut für Physikalische und Theoretische Chemie, Universität Leipzig, Linnéstr. 2, 04103 Leipzig, Germany

Jae Hyo Han
Center for Nanomedicine, Institute for Basic Science (IBS), Seoul 03722, Republic of Korea
Yonsei-IBS Institute, Yonsei University, Seoul 03722, Republic of Korea
Department of Chemistry, Yonsei University, Seoul 03722, Republic of Korea

Jinwoo Cheon
Center for Nanomedicine, Institute for Basic Science (IBS), Seoul 03722, Republic of Korea
Yonsei-IBS Institute, Yonsei University, Seoul 03722, Republic of Korea
Department of Chemistry, Yonsei University, Seoul 03722, Republic of Korea

Thomas Heine
Chair of Theoretical Chemistry, TU Dresden, Mommsenstr. 13, 01062 Dresden
Helmholtz-Zentrum Dresden-Rossendorf, Abteilung Ressourcenökologie, Forschungsstelle Leipzig, Permoserstr. 15, 04318 Leipzig, Germany
Wilhelm-Ostwald-Institut für Physikalische und Theoretische Chemie, Universität Leipzig, Linnéstr. 2, 04103 Leipzig, Germany
thomas.heine@tu-dresden.de

**Abstract**

Two-dimensional crystals, single sheets of layered materials, often show distinct properties desired for optoelectronic applications, such as larger and direct band gaps, valley- and spin-orbit effects. Being atomically thin, the low amount of material is a bottleneck in photophysical and photochemical applications. Here, we propose the formation of stacks of two-dimensional crystals intercalated with small surfactant molecules. We show, using first principles calculations, that already very short surfactant methyl amine electronically decouples the layers. We demonstrate the indirect-direct band gap transition characteristic for Group 6 transition metal dichalcogenides experimentally by observing the emergence of a strong photoluminescence signal for ethoxide-intercalated $WSe_2$ and $MoSe_2$ multilayered nanoparticles with lateral size of about 10 nm and beyond. The proposed hybrid materials offer the highest possible density of the two-dimensional crystals with electronic properties typical for monolayers. Variation of the surfactant's chemical potential allows fine-tuning of electronic properties and potentially elimination of trap states caused by defects.

**Article**

The field of two-dimensional crystals (2DC) has been fuelled by applications in nanoelectronics[1,2] and photocatalysis.[3] Among the plethora of 2DC,[4–6] we find a number of materials that have the desired optical, photophysical, and photocatalytic properties for these applications, while their 3D counterparts, which share the same stoichiometry and chemical bonding pattern, do not. Examples include Dirac cones at the Fermi level in graphene that are absent in graphite,[7] larger and direct band gaps in Group 6 transition metal dichalcogenides (TMDCs)[8–11] and antimonene[12] the semiconductor-metal transition between single layer and bulk of $GeP_3$[13] or $PtSe_2$,[14] and large spin-orbit splitting due to reduction of symmetry in monolayers (MLs) of 2H TMDCs.[15–17] Yet, a single or few atoms thin ML is not sufficient for practical photovoltaic or photocatalytic applications, as the tiny amount of material does not provide the sufficient absorbance to the incoming light.

Recently, Wang *et al*.[18] reported so-called monolayer atomic crystal molecular superlattices, layered materials intercalated with molecular species which increase the interlayer distance to an extent that the layers are electronically decoupled and get the properties as their corresponding MLs. For example, a heterocrystal made of black phosphorus, which is electrochemically intercalated with cetyl-trimethylammonium bromide, shows the excellent conductive properties of the black phosphorus ML. The same approach was then successfully applied to a series of metal chalcogenides, including $MoS_2$ and $WSe_2$, showing their transition to the direct band, and further materials such as SnSe, GaS, $NbSe_2$, $In_2Se_3$ and $Bi_2Se_3$. While this suggests a technology that is applicable in general, it is important to note that the proposed electrochemical intercalation enlarges the interlayer distance by at least 9 Å. This poses the question, if the same result could be achieved at smaller interlayer distances, and if it is applicable to milder intercalation approaches, in particular to surfactant intercalation in liquid phase.

Indeed, such surfactant intercalated layered materials occur during the liquid exfoliation process and during colloidal synthesis in presence of surfactants.[19–21] We have already demonstrated that surfactant-intercalated layered nanoparticles $TiS_2$, $WSe_2$, $ZrS_2$, $NbS_2$, and $MoS_2$ are stable, and that the interlayer distance depends linearly on the chain length of the surfactant.[20]

Also, it is interesting to know if the surfactant's head group can alter the electronic properties of the 2DC layers. Tuning properties of layered materials via intercalating charged species is known for a long time.[22,23] Among neutral intercalates, polymers,[24] heterolayers,[25] and molecules have been studied.[26–28] Förster *et al.* recently showed that point defects in TMDCs can be healed by the presence of surfactants.[29]

Here, we show that electronic decoupling of the layers is also achieved in the surfactant-based intercalation approach, and that electronic layer separation is achieved upon lattice enlargement by ~4 Å only. This value is obtained for surfactants with the shortest side chain, such as methylamine, $MeNH_2$, (**Figure 1 a**), and we show that the electronic structure of the 2DC can be fine-tuned by the surfactant's head group.

For our calculations, we have selected four examples with pronounced quantum confinement effects: $1T$-$TiS_2$ is a semi-metal with high conductivity in the layered bulk.[30,31] $2H$-$MoS_2$, which shows the transition from indirect to direct band gap when going from bulk to a monolayer[32–34] and giant spin-orbit splitting of ~150 meV at the valence band maximum of the monolayer, which is absent in the bulk.[15,17,35] Isoelectronic 2H-WSe$_2$ shows similar effects as 2H-MoS$_2$, but more pronounced spin-orbit splitting of ~450 meV.[15,35] Finally, 1T-PtSe$_2$ [36,37] shows extreme quantum confinement with metal-semiconductor transition when going from bulk to the monolayer.[38,39] We restrict our calculations to one intercalate, methylamine (MeNH$_2$), which is the shortest possible surfactant with an amine group.

To complement this study, we have carried out photoluminescence experiments (see Supporting Information (SI) for details), in which, we have intercalated WSe$_2$ with slightly longer short-chained ethoxide molecules and show the recovery of the strong photoluminescence signal that is characteristic for WSe$_2$ monolayer.

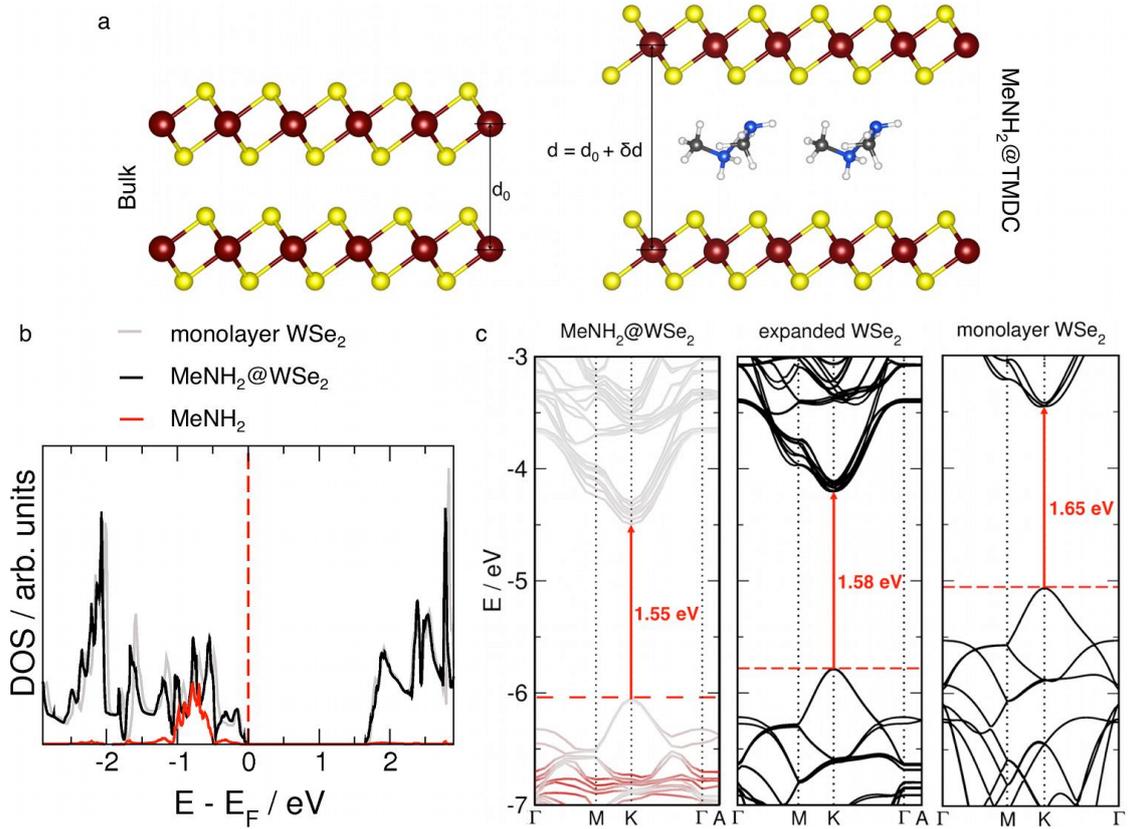

**Figure 1** – (a) Scheme of TMDC bulk (left) and intercalated bulk with MeNH$_2$ (heterostructure; right). $d_0$ - interlayer distance in perfect bulk between metal cores, d - interlayer distance in the heterostructure, δd - increase in the interlayer distance of the bulk due to intercalates. (b) Density-of-states (DOS) around the Fermi-level, $E_F$, of intercalated bulk MeNH$_2$@WSe$_2$, with projections of the TMDC layers (black) and the intercalate (red), compared with the reference (gray) of the expanded bulk structure (cf. Figure S3). (c) Band structures: MeNH$_2$@WSe$_2$ - intercalated TMDC bulk (heterostructure); WSe$_2$ - host layers as in the intercalated TMDC system (expanded bulk); monolayer WSe$_2$ - perfect TMDC monolayer. Horizontal dashed lines indicate the Fermi levels with respect to the vacuum. Gray - band contribution from the TMDC layers, red - band contribution from the intercalate molecules.

Figure 1 a shows WSe$_2$, as exemplary layered crystal, with and without MeNH$_2$ intercalation, which expands the interstitial space by δd upon formation of heterostructures (HS). The same setup is utilised in all other TMDC calculations. The structural properties of bulk and monolayer TMDCs are shown in Table S1 in SI, while the structural properties of the HS are given in Table S2 and the structures are shown in Figure S1. Upon intercalation of MeNH$_2$, the host TMDC bond lengths and angles change very little (maximum differences of 2%), supporting weak binding of the intercalate molecules to the basal planes of TMDC. The interlayer expansion values (δd, see Figure 1) due to the formation of HS are given in Table S3 for all the systems considered here. Depending on the TMDC, MeNH$_2$ molecules rearrange themselves such that δd varies between 3.4 Å (WSe$_2$) and 4.8 Å (MoS$_2$). In contrast to the results of Wang et al.,[18] the chain length of our intercalate molecules is very small, hence they cause very little strain on the layers. In case of longer chains, e.g., TiS$_2$ intercalated with alkyl-amines (propyl … hexyl), the molecules arrange themselves perpendicularly with respect to the crystal planes, as revealed by an analysis based on TEM and simulation data,[20,40] and can be rationalised by the rather small van-der-Waals forces in between the short-chained surfactant molecules.

It should be noted that the choice of the surfactant molecule in the present studies, MeNH$_2$, serves as proof of concept and might not be the most optimal for all the selected TMDC. The choice of the surfactant should be motivated by practical questions that arise in experiment. For this purpose, the relative chemical potentials of layered materials and the surfactant should be considered. Nevertheless, whatever surfactant is selected, the electronic structure of the individual layers will be decoupled already for very short ones.

By topological analysis, charge transfer study and by the analysis of the band structures, we show that the layers are essentially unaffected by the intercalate itself. The main effect on the electronic structure is due to the enlargement of the interstitial space. Hence, this choice needs to be motivated by practical questions, e.g., quantitative intercalation.

The weak interaction of MeNH$_2$ with TMDCs is supported by the energetic analysis (Table S3), with the expansion energies being close in values to the corresponding exfoliation energies. The Hirshfeld charge analysis revealed that only in the case of TiS$_2$, methylamine acts as an electron donor. The electrons are equally donated to both Ti and S atoms. In the other three cases, the chalcogen atoms attract charge, particularly strong for the two selenides, as shown by charge transfer of ~0.12 e$^-$ per MX$_2$. Further, Bader analyses of the electronic density did not reveal any bond critical points between the MeNH$_2$ and TMDC layers (small positive values of the Laplacian of the density ($\Delta\rho$); see Table S4).

In Figure 1 b and S2, we compare the density of states (DOS) of the heterostructures with these of the expanded bulks (see Figure S3 for definition of the expanded bulk). For better comparison, we shifted the DOS of HS to overlap the characteristic features of the DOS of the expanded bulks. In this way, we could distinguish two cases: i) in TiS$_2$ and WSe$_2$ the Fermi level of the expanded bulks and HS are the same, while ii) for MoS$_2$ and PtSe$_2$, the Fermi level of HS is shifted up in the energy.

On the other hand, the band structures (see Figure 1 c and Figures S4-7) indicate the differences in the chemical potentials of TMDC and the MeNH$_2$ intercalates. In the exemplary case of WSe$_2$ (cf. Figure 1 c), the chemical potential of the MeNH$_2$ is lower than that of the perfect ML and nearly the same as in the expanded WSe$_2$, resulting in no intercalate states in the top of the valence band. The opposite situation is found in the case of MoS$_2$ and PtSe$_2$, where the intercalate donates midgap states. In the case of TiS$_2$, the intercalate molecules open a small band gap (of about 30 meV) most likely due to the charge transfer from intercalate to the TMDC basal plane. The 3$d$ orbitals of Ti are not screened as compared to the elements of the 5th and 6th period, what makes them more delocalized and allows partial overlap with 2$p$ orbitals of the amine group.

Moreover, electronic structure analysis of $MoS_2$ and $WSe_2$ shows that the states close to the Fermi level, at the Γ point, are dominated by the $p_z$ chalcogen atoms; therefore, they are the most affected by the charge transfer and are lowered in energy in the intercalated systems. The states at the K points are dominated by the in-plane metal *d*-orbitals and remain mostly unaffected, thus, the direct band gap of the single layers is maintained (see Figures S8-10).

In $MoS_2$, we predict a separation of 4.8 Å, which causes the layers to behave almost identically to the monolayers, that is, the band gap is direct with a value of 1.8 eV. For $MeNH_2$@$WSe_2$, we calculated a band gap of 1.55 eV, which is about 100 meV smaller than in the perfect monolayer. This is due to the smaller separation (δd = 3.4 Å), which does not decouple the layers completely. As observed in our experiment, it requires an additional member in the alkyl chain to recover the perfect $WSe_2$ ML in the 3D assembly (see discussion below). Bulk $PtSe_2$ is semimetallic, while the monolayers are indirect band gap semiconductors. The perfect ML and HS show indirect band gaps of about 1.3 eV. This value converges for δd of at least 3 Å. For comparison, in Figure S11 and S12, we have shown the effect of the interlayer separation (δd) on the change of the band gap for δd = 0–3 Å. These calculations were performed without the intercalate molecules, therefore, showing purely the effect of the interlayer expansion.

The experimental procedure and the transmission electron microscopy (TEM) images are shown in **Figure 2 a**. The initial interlayer distance of pristine $WSe_2$ is 6.5 Å, but after intercalation of meth- and ethoxide, it increases to 8.0 Å and 13.1 Å, respectively, according to the TEM analysis (Figure 2 b-d, for details please see Figure S13 and Ref. [20]). In order to determine whether the intercalation of small alkoxide molecules can result in a superlattice structure with electronically decoupled monolayers in a multilayer stack, we have examined the photoluminescence (PL) spectra of the obtained colloidal solutions. Figure 2 e shows that both the pristine $WSe_2$ and the methoxide-intercalated $WSe_2$ exhibited no PL signals, whereas ethoxide-intercalated WSe$_2$ showed a prominent PL emission. Similar behaviour was seen in MoSe$_2$ nanoflakes. The blue-shifted broad PL emission from the literature value of 1.62 eV for CVD grown or mechanically exfoliated single-layer WSe$_2$ is attributed by the lateral quantum confinement effect with size heterogeneity in ensemble (see Figure S14).

Due to low-lying d$_{z^2}$ orbitals in Group 6 TMDCs, the choice of the intercalate by colloidal synthesis is experimentally limited and it is rather difficult to intercalate the linear alkylamines.[41,42] Alternatively, a stronger base, such as alkoxides, can be used, for which a successful intercalation has been previously reported with adequate amount.[40]

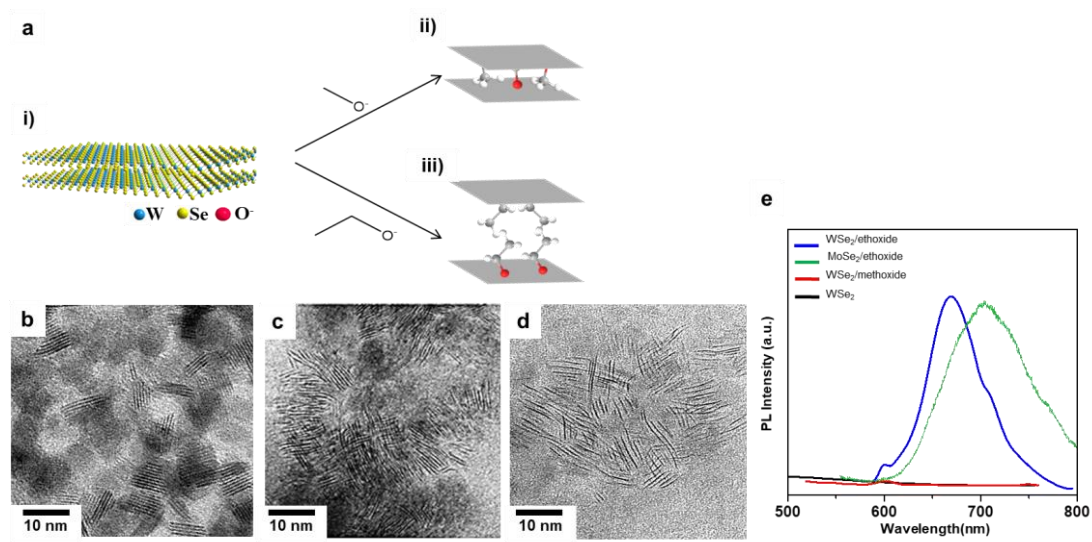

**Figure 2.** Interlayer expansion of WSe$_2$ quantum dots with alkoxides molecules as intercalates, with different chain lengths. a) Schematics of the interlayer distance change in WSe$_2$ by intercalating alkoxides. (b-d) Side view of layers in the TEM images: (b) before intercalation and after intercalation of (c) methoxide and (d) ethoxide. (e) Photoluminescence spectra observed from colloidal solutions of pristine WSe$_2$ (black), methoxide (red) and ethoxide intercalated WSe$_2$ (blue) and MoSe$_2$ (green) multi-layered quantum dots.

Our experimental results on alkoxide-intercalated WSe$_2$ show that the observed expansion of the interlayer gap results in a monolayer arrangement for methoxide composites, whereas we observe an interdigitated bilayer arrangement for ethoxide systems. The small gap expansion obtained by the intercalation of methoxide ($\delta d$ = 1.5 Å) is not sufficient to decouple the system, in agreement with our theoretical results. On the other hand, a larger gap expansion in the ethoxide-intercalated WSe$_2$ ($\delta d$ = 6.6 Å) is definitely able to change the lowest allowed transition from an indirect to a direct band gap. This agrees with our theoretical predictions.

It is important to mention that our theoretical analysis does not account for lateral quantum confinement, which has been well-studied for TiS$_2$ [43] and WSe$_2$.[44] To keep our conclusions sound, the TMDC nanoparticles should have a lateral diameter of at least 10 nm. The lateral confinement offers another option to tune the electronic properties in the multi-layer stack, and the electronic decoupling is directly transferable to larger nanoflakes.

In summary, we have investigated the formation of 3D assemblies of monolayer TMDCs by intercalation of small intercalate molecules into bulk forms. The molecules are large enough to electronically decouple bulk layers, the intercalates interact weakly with the TMDC basal planes, therefore, the intercalated system exhibits electronic properties of the corresponding TMDC monolayers. This is confirmed by our photoluminescence experiments on the example of the ethoxide-intercalated bulk WSe$_2$, which recovers the direct band gap of the WSe$_2$ monolayer. The distance when the layers get electronically decoupled is, depending on the system, between 1 and 3 Å.

Differences in the chemical potentials of intercalates and TMDCs results in heterostructures with electronic properties of respective monolayers (WSe$_2$, TiS$_2$) or in cases, where intercalates contribute states that affect the Fermi level (MoS$_2$, PtSe$_2$). The latter could be beneficial in defect healing.

The formation of 3D assembly of 2D monolayers should be more promising than single layers for optoelectronic, photovoltaic, and photocatalytic applications, where increased cross section of photoactive monolayers is desirable, and here it is offered by stacking of many of such layers by colloidal synthesis. We also note here that the inner layers of these heterostructures are chemically stabilized, which offers an alternative to hBN encapsulation. In future, 3D integration of electronic devices made of single-layer transistors could become feasible using the surfactant intercalation approach.

**Supporting Information**

Supporting Information is available from the Wiley Online Library or from the author.

Acknowledgements

The Deutsche Forschungsgemeinschaft (HE 3543/35-1) and AOARD 134142 (contract no. FA2386-14-1-0014) are gratefully acknowledged. This work was supported by the Institute for Basic Science (IBS-R026-D1).

The authors thank Dr. Thomas Brumme (Leipzig University) for fruitful discussions, as well as Dr. Augusto Faria Oliveira (Jacobs University Bremen), Dr. Marc Raupach and Michal Handzlik (SCM, Amsterdam) for their technical support.

We acknowledge ZIH Dresden for computational support.

**Author contributions**

RK, AK and TH contributed in the calculations and analysis of theoretical data; JHH and JC contributed in the experiments and analysis of measure data; all authors contributed in writing this paper.

**Optoelectronic properties of 2D crystals are distinct from their bulk counterparts, and advantageous for applications.** Band gaps are larger, sometimes direct, well-suited for optoelectronic and photocatalytic applications. However, being atomically thin, low density is a bottleneck for quantitative performance. The proposed intercalation of layered materials with short surfactants yields electronically decoupled layers with similar properties as monolayers, but higher cross-section.

**Keyword** 2D materials, Heterostructures, Intercalates, Optoelectronics, Photoluminescence

Roman Kempt, Agnieszka Kuc, Jae Hyo Han, Jinwoo Cheon, Thomas Heine*

**Two dimensional crystals in three dimensions**

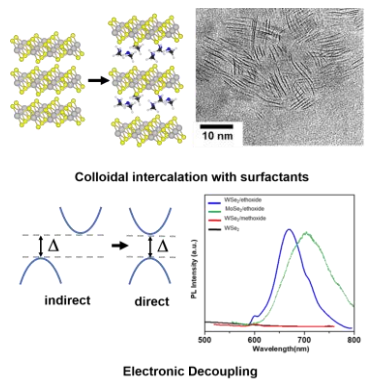